\documentclass[conference]{IEEEtran}
\IEEEoverridecommandlockouts
\usepackage{cite}
\usepackage{amsmath,amssymb,amsfonts}
\usepackage{optidef}
\usepackage{algorithmic}
\usepackage{graphicx}
\usepackage{textcomp}
\usepackage{xcolor}
\def\BibTeX{{\rm B\kern-.05em{\sc i\kern-.025em b}\kern-.08em
    T\kern-.1667em\lower.7ex\hbox{E}\kern-.125emX}}

\author{
    \IEEEauthorblockN{Lakshya Jagadish\IEEEauthorrefmark{1}, Banashree Sarma\IEEEauthorrefmark{1}, R. Manivasakan\IEEEauthorrefmark{1}}
    \IEEEauthorblockA{\IEEEauthorrefmark{1}Indian Institute of Technology Madras, Chennai, India
    \\\{ee19b035@smail.iitm.ac.in, ee18d036@smail.iitm.ac.in, rmani@ee.iitm.ac.in\}}
}

\title{Multi Agent DeepRL based Joint Power and Subchannel Allocation in IAB networks\\
}

\begin{document}

\maketitle

\begin{abstract}
Integrated Access and Backhauling (IAB) is a viable approach for meeting the unprecedented need for higher data rates of future generations, acting as a cost-effective alternative to dense fiber-wired links. The design of such networks with constraints usually results in an optimization problem of non-convex and combinatorial nature. Under those situations, it is challenging to obtain an optimal strategy for the joint Subchannel Allocation and Power Allocation (SAPA) problem. In this paper, we develop a multi-agent Deep Reinforcement Learning (DeepRL) based framework for joint optimization of power and subchannel allocation in an IAB network to maximize the downlink data rate. SAPA using DDQN (Double Deep Q-Learning Network) can handle computationally expensive problems with huge action spaces associated with multiple users and nodes. Unlike the conventional methods such as game theory, fractional programming, and convex optimization, which in practice demand more and more accurate network information, the multi-agent DeepRL approach requires less environment network information. Simulation results show the proposed scheme's promising performance when compared with baseline (Deep Q-Learning Network and Random) schemes.
\end{abstract}

\begin{IEEEkeywords}
Integrated Access and Backhaul, Deep Reinforcement Learning, Power Allocation, Subchannel Allocation
\end{IEEEkeywords}

\section{Introduction}
With data traffic growing exponentially and the demand for more capacity, mm-Wave seems to be a viable solution. But owing to the higher path loss and attenuation of mm-Wave, cell size must be reduced~\cite{6732923}. While network densification may be a viable option for mm-Wave which also meets the goal of an ultra-dense network, setting up wired backhaul links is not economically feasible. In the mm-wave networks using wireless backhauling IAB nodes can be installed fairly easily compared to optical fiber networks. Furthermore, mm-wave wireless backhaul can boost network capacity and spectrum efficiency. Since the same spectrum is used for access and backhaul links, an efficient resource allocation is warranted for such a system. 

Although the resource allocation problem for an IAB system has been studied extensively, most previous work has used complex conventional ways to solve it where the full or partial CSI (Channel State Information) is required. Much of the previous work deals with solving optimization problems with high computational complexity. Because of the unpredictability of future wireless environments, rule-based decision-making that selects decisions directly from training may not be ideal. As a result, it may not be effective to design a priori cost functions and then solve optimal control problems in real time.
DeepRL is being prominently used for solving such problems in 5G\cite{3104322}. The approaches can be centralized or decentralized.
Since the information of all the IAB nodes should be reported to the central controller for solving the resource allocation optimization problem, the transmission overhead is large. It grows dramatically with the size of the network, which prevents these methods from scaling to large networks. Therefore, in this paper, we focus on decentralized resource allocation approaches.

Advancements in the field of Reinforcement Learning (RL) have provided the opportunity of finding effective solutions to such resource allocation and optimization problems using Q-Learning, deep Q-learning (DQN), and double deep Q-learning (DDQN). Q-learning is the most basic architecture. However, it requires the maintenance of huge tables and leads to slower convergence in large state-action spaces. DQN uses a deep neural network (DNN) as a functional approximation for the Q-Table of the state-action pairs instead of storing Q-values for each state-action pair. This can lead to faster convergence to the optimal solution but can lead to overfitting, which can depreciate the performance of the trained model. DDQN decomposes the max of objective function operation (downlink data rate in our case) for the given state in the target network into action selection and evaluation. Therefore the greedy policy is evaluated according to the online network (evaluation), using the target network to estimate its value (selection). The weights of the target network are periodically updated according to the weights of the online network.

To overcome the drawbacks of the traditional totally centralized and distributed deep reinforcement learning-based resource allocation approaches, we propose a multi-agent deep Q-Learning algorithm with decentralized learning and centralized execution to solve the formulated optimization problem. This problem is formulated as a mixed integer non-linear optimization problem, intending to maximize the downlink sum-rate throughput.

\subsection{Related Work}
Many studies have been done on resource allocation in wireless backhaul networks.
~\cite{9625522} dynamically allocates the spectrum using an auction-based design of the system. The main limitations of these conventional approaches are computational complexities apart from the need for complete channel state information, which might not be feasible for an ultra-dense network. The downlink sum rate or network capacity is optimized in~\cite{9133540}and~\cite{9217104}. ~\cite{DAYAN2008185} talks about penalizing the agent when it does not output an optimal state. Studies ~\cite{10027135},~\cite{10065879},~\cite{10067029},~\cite{9864179},~\cite{10049013},~\cite{9725256},~\cite{9086877} and~\cite{9473755} abandon the conventional optimization framework and use different flavors of DeepRL. We use this idea to give "negative rewards" or "penalties" to deter the agent from making such decisions in the future. In this paper, we perform the proposed \textit{SAPA-DDQN} overcoming the constraints of huge action space and centralized overhead mechanisms. This network can be trained offline, hence is suitable for practical deployment. The model does not require much retraining unless there are variations in the positions of the nodes and the topology of the network. To our best knowledge, our paper makes the first attempt to use the DBS and IABs as agents for training the DDQN, in a centralized and decentralized manner.

The paper is organized as follows: In Section \ref{sysmodel}, we elaborate on the system model. In Section \ref{probform} we delve into how the optimization problem we aim to solve. Section \ref{madrl} explains in detail the machine learning model we employ to achieve a solution to the SAPA problem. Finally, we present the model results in Section \ref{sim} and conclude the paper in Section \ref{conc}.

\section{SYSTEM MODEL}\label{sysmodel}
\begin{figure}
    \centering
    \includegraphics[scale=0.30]{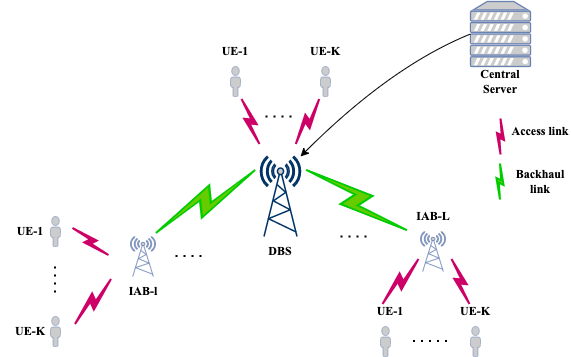}
    \caption{A two-tier IAB network}
    \label{system}
\end{figure}

A two-tier IAB network is considered, as shown in Figure \ref{system}, consisting of one IAB donor at the center which is connected to the core network through fiber. $L$ IAB nodes are uniformly distributed around the donor base station (DBS) within a coverage of radius $R$. Let $\mathcal{L}=\{0 \leq l \leq L\}$ denote the set of all base stations (BS) (DBS and IAB nodes) where $0$ represents the DBS while $1 \leq l \leq L$ denotes the $L$ IAB nodes. IAB nodes are assumed to communicate in Full Duplex (FD) mode resulting in undesirable self-interference. The terms nodes and BS has been used interchangeably. The users are associated with their nearest BS, and a user can connect to at most one BS. Each BS can connect to $K$ users forming the set $\mathcal{K}_l=\{1\leq k\leq K\}$ corresponding to the $l^{th}$ base station for the sake of simulations.
Here we have considered integrated spectrum allocation, where the same spectrum is used by the DBS and each IAB node. The total bandwidth has been partitioned into $N$ orthogonal subchannels, denoted by $\mathcal{N}=\{1\leq n\leq N\}$. We have assumed each BS applies OFDMA to its users, i.e., each subchannel is allocated to only one user at a time (access) apart from its (self) backhaul.

For simplicity, we assume flat fading within a subchannel and channel fading is independent across the $N$ subchannels. We denote the downlink backhaul channel gain between the donor and $l^{th}$ IAB node in the $n^{th}$ subchannel as
\begin{equation}
    g_{0,l}^{n} = h_{0,l}^{n}\alpha_{0,l},\ \ l \neq 0
\end{equation}
Here, $\alpha_{0,l}$ represents the large-scale fading component, including path loss and log-normal shadowing. $h_{0,l}$ represents the Rayleigh fading. Similarly, let $g_{l,k_{l}}^{n}$ denote the downlink access channel gain, where $l=0$ denotes the donor-to-user access link whereas the access link between an IAB node and its users are represented by $l>0$.

Since we are also solving the problem of efficient power allocation, we define the possible power levels of the IAB donor shared between the access and backhaul links as $\mathcal{P}_0 = \{P_1, P_2,..., P_{T_1}\}$ where the transmitting power $P_{DBS}$ is discretized into $T_1$ levels and the set of power levels is therefore denoted by $\mathcal{P}_0=\{0,\frac{P_{DBS}}{T_1-1},\frac{2P_{DBS}}{T_1-1},...,P_{DBS}\}$ and $p_j \in \mathcal{P}$. Hence, the power allocation vector from the DBS ($l=0$) to its associated IAB nodes ($l>0$) is denoted by $\mathit{P}^{DN} = [{p}^{n_1}_{0,1}, {p}^{n_2}_{0,2},...,{p}^{n_L}_{0,L}]$, where ${p}^{n_l}_{0,l} \in \mathcal{P}_0$ and $n_l$ is the allocated subchannel. Similarly, the power allocation vector from the DBS to its associated $K$ users is denoted by $\mathit{P}^{DU} = [{p}^{n_1}_{0,1}, {p}^{n_2}_{0,2},...,{p}^{n_K}_{0,K}]$, where ${P}^{n_k}_{0,k} \in \mathcal{P}_0$ and $n_k$ is the allocated subchannel associated with the access link. The power ($P_{IAB}$) of each of the $L$ IAB nodes are divided similarly to $T_2$ power levels. We denote the possible power allocations by $\mathcal{P}_L = \{P_1, P_2,...,P_{T_2}\}= \{0,\frac{P_{IAB}}{T_2-1},\frac{2P_{IAB}}{T_2-1},...,P_{IAB}\}$. The power allocation vector from the $l^{th}$ IAB node to its associated UEs is denoted by $\mathit{P}^{NU} = [{p}^{n_1}_{l,1}, {p}^{n_2}_{l,2},...,{p}^{n_K}_{l,K}]$, where ${p}^{n_k}_{l,k} \in \mathcal{P}_L$ and $n_k$ is the allocated subchannel between the IAB node and the associated UE. We finally describe the overall power allocation vector as the concatenation of the three vectors $P^{DU}$, $P^{DN}$ and $P^{NU}$.
\begin{equation}
    P = [P^{DU}; P^{DN}; P^{NU}]
\end{equation}

Together with co-channel interference, since IAB nodes are assumed to be in FD (Full Duplex) mode, self-interference will also be experienced, i.e., the BSs on the same sub-band (DBS and IAB nodes) also affect the transmission. Taking these interferences into account, the received Signal-to-interference-plus-noise ratio (SINR) of the $l^{th}$ IAB node in $n^{th}$ subchannel can be written as:

\begin{equation}\label{2}
    \gamma_{0,l}^{n}=\frac{p_{0,l}^{n}g_{0,l}^{n}}{\sum_{j\neq l}^{L}\sum_{k=1}^{K}{p_{j,k_{j}}^{n}g_{j,k_{j}}+\beta_{l}p_{l,k_l}g_{l,k_l}+\textit{Noise}}}
\end{equation}

The numerator corresponds to the signal power in consideration. The first term of the denominator is the signal due to the interferers, the second term corresponds to self-interference. $p_{j,k_{j}}^{n}$ is the transmission power of the $j^{th}$ interfering node. $\beta_l$ denotes the self-interference cancellation ability and $0\leq\beta_l\leq1$. $\beta_l=0$ means full self interference cancellation and $\beta_l$=1 means no self interference cancellation.
 
Let $k_l$ denote user $k$ associated with the BS $l$. In a similar way, the SINR at the receiver of $k_l^{th}$ user associated with the $l^{th}$ IAB is:
 \begin{equation}
\gamma_{l,k_l}^{n}=\frac{p_{l,k_l}^{n}g_{l,k_l}^{n}}{\sum_{j\neq l}^{L}\sum_{m=1}^{K}{p_{j,k_{m}}^{n}g_{j,l_{m}}+\beta_l\Hat{G}_{l,k}+\textit{Noise}}}
 \end{equation}
 where
 \begin{equation}
     \Hat{G}_{l,k} = \sum_{j=l}^{L}P_{j}^{DN}g_{l,k}^{n} + \sum_{m=1}^{K}P_{m}^{DU}g_{l,k}^{n}
 \end{equation}
 which stands for the interference introduced by the DBS to the $k^{th}$ UE associated with the $l^{th}$ IAB node. Here we are indexing the $j^{th}$ item from the vector $P^{DN}$ and the $m^{th}$ item from the vector $P^{DU}$.
 The SINR of the $k_0^{th}$ user associated with the DBS can be expressed as:
 \begin{equation}
    \gamma_{0,k_0}^{n}=\frac{p_{0,k}^{n}g_{0,k_0}^{n}}{\sum_{j=1}^{L}\sum_{k=1}^{K}{p_{j,k_{j}}^{n}g_{j,k_{j}}+\textit{Noise}}}
 \end{equation}
The data rate of the $k_0^{th}$ user belonging to the network, associated with the DBS, can be given as $R_{k_0}=\sum_{n=1}^{N} \mathbf{x}_{0, n}\log_2 (1+\gamma_{0,k_0}^n)
$
Here we introduce the subchannel allocation vector $\mathbf{x}$ which denotes the allocation of subchannels to the users from the DBS and IAB nodes as\\$x_{l,n} = 
    \begin{cases}
        1 & \text{if}\ n^{th}\ \text{subchannel is allocated to the}\ k^{th}\ \text{user}\\
        0 & \text{otherwise}
    \end{cases}$ where $l \in \mathcal{L}$.
$x_l = [x_{l,0}, x_{l,1}, ..., x_{l,N}]$, combining this element for the multiple subchannels we obtain the subchannel allocation vector associated with the DBS or IABs. Similarly, $x_{0l} = [x_{0l, 0}, x_{0l, 1}, ..., x_{0l,N}]$ denotes the allocation of subchannels from the DBS to the IABs as follows where $l \in \mathcal{L}-\{0\}$
Finally, we obtain the subchannel allocation matrix as follows
\begin{equation}
    \mathbf{x} = [x_0, x_1, ..., x_L, x_{01}, x_{02}, ..., x_{0L}]
\end{equation}
Similarly, the data rate of the $k_l^{th}$ user, associated with the $l^{th}$ IAB node is given as:
\begin{equation}\label{7}
R_{k_l}=\min\{\sum_{n=1}^{N}\mathbf{x}_{0l, n}\log_2 (1+\gamma_{0,l}^n),\sum_{n=1}^{N}\mathbf{x}_{l, n}\log_2 (1+\gamma_{l,k_l}^n)\}
\end{equation}
where $\forall\ {l}\in \mathcal{L}-\{0\}$.
Here, the first term refers to the log-sum rate of the backhaul links and the second term refers to the log-sum rate of the access links.

\section{PROBLEM FORMULATION}\label{probform}
We are interested in maximizing the sum rate of the users as the objective function. The joint subchannel allocation and power allocation (SAPA) problem at a discretized time stamp $t$ is formulated as
\begin{maxi*}|l|
{\mathbf{x},P}{\sum_{k\in \mathcal{K}_l}{\sum_{l\in \mathcal{L}}{R_{k_l}}}}
{}{}
\addConstraint{\sum_{l\in \mathcal{L}}p_{0,l}^n + \sum_{k\in \mathcal{K}_l}p_{0,k}^n\leq P_{DBS}}
\addConstraint{\sum_{k \in \mathcal{K}_l}{P_{l, k_l}^n \leq P_{IAB}}, \; \forall\ {l} \in \mathcal{L}-\{0\}}{}
\end{maxi*}
In the context of the DeepRL problem, $\mathbf{x}$ implies the subchannel distribution matrix for the IAB nodes and DBS, and $P$ implies the power distribution vector for each of the IAB nodes and DBS. These two formulate the action space of our problem, which we aim to find for an optimized resource allocation. The constraint of no two nodes in the same tier sharing the same subchannel applies here as well.
We consider two special cases. In the first case, the objective is to maximize the downlink data rate for all time stamps $t$ by maintaining stationary UE positions. In the second case, the UEs are perturbed according to certain constraints. We observe the differences and report the same in the later sections.

\section{MULTI AGENT DQN FOR RESOURCE ALLOCATION IN IAB}\label{madrl}
In this section, we illustrate the use of MADRL (Multi Agent Deep Reinforcement Learning) to solve the aforementioned NP-hard optimization problem in the context of distributed execution and centralized training. We first understand the basics of Deep reinforcement learning and then, model the problem as a Markov Decision Process (MDP) and define its agents, states, actions, and rewards.

\subsection{Deep Reinforcement Learning Basics}
A reinforcement learning\cite{8792117} agent learns its best policy from observing the rewards of trial-and-error interactions with its environment over time. Let $\mathcal{S}$ denote a set of possible states and $\mathcal{A}$ denote a discrete set of actions. The state $s \in \mathcal{S} $ is a tuple of the environment’s features that are relevant to the problem at hand and it describes the agent’s relation with its environment. Assuming discrete time steps, the agent observes the state of its environment, $s_t \in \mathcal{S}$ at discrete time step $t$. It then takes an action $a_t \in \mathcal{A} $ according to a certain policy $\pi$. The policy $\pi(s, a)$ is the probability of taking action $a$ conditioned on the current state $s$. The policy function must satisfy $\sum_{a\in A}\pi(s,a)=1$. Once the agent takes an action $a_t$, its environment moves from the current state $s_t$ to the next state $s_{t+1}$. As a result of this transition, the agent gets a reward $r_{t+1}$ that characterizes its benefit from taking action $a_t$ at state $s_t$. This scheme forms an experience at time $t+1$, hereby defined as $e_{t+1} =(s_t,a_t,r_{t+1},s_{t+1})$, which describes an interaction with the environment\cite{mnih2015human}. For an agent to make decisions aligned with the learning objective, the reward obtained for each state-action pair must align with the learning process and objective. 

\subsection{The MDP model}
The dynamic environment in this scenario is the IAB network, which has a donor in the center and $L$ IAB nodes surrounding it.
The donor and IAB nodes act as learning agents, and each agent learns an optimal policy to maximize the system's sum rate through trial-and-error interactions with the environment. Let $\mathcal{P}_X$ denote the set of probability distributions over the set $X$. In our case, $X$ would be $\mathcal{S}$, which denotes the state space. In MADRL, the $j^{th}$ agent is represented as a tuple $\langle S_j, A_j, R_j, P_j, \gamma\rangle$, where $S_j$ is the state, $A_j$ is the action, $R_j:\mathcal{S}\times\mathcal{A}\xrightarrow{}\mathbb{R}$ is the reward function, and $P_j:\mathcal{S}\times\mathcal{A}\xrightarrow{}\mathcal{P}_\mathcal{S}$ is the state transition probability function of the $j^{th}$ agent. $0\le\gamma\le 1$ is a discount factor on the summed sequence of rewards. The learners (or agents) receive $S_j, A_j,$ and $\gamma$ as input. The agent always occupies a single state $x$ of the MDP $\mathbb{M}$. The agent is told this state and must choose an action $a$. It then receives an immediate reward $r \sim R_j(s,a)$ and is transported to a next state $\tilde{s} \sim P_j(s,a)$. The first state occupied by the agent may be chosen arbitrarily, in our case we use a random assignment technique. The goal of each of the agents in the environment is to maximize the rewards obtained in the shortest time possible. The states, actions, and rewards of each agent are discussed below in case of our problem formulation is explained below:

\textbf{States}: As we consider the MARL (Multi-Agent Reinforcement Learning) framework in the network, each agent $j$ (i.e., donor or IAB node) only observes the network state of those devices which are associated with it. Thus, the state of $l^{th}$ agent at time $t$ can be given as $s_{l,t}=\{\{s_{k_l}\}\}$, where $s_{k_l}=1$ if the rate of $k_l^{th}$ user associated with the $l^{th}$ IAB node is greater than the threshold, $R_{th}$.

\textbf{Actions}: On observing the current state, each agent takes an action $a_j^t \in {A_j}$ based on the decision policy $\pi$. The action space $A_j$ of $j^{th}$ agent contains all possible resource allocation choices, which is the combination of the selection of a subchannel and the corresponding transmission power. Therefore, the action chosen by the $j^{th}$ agent can be expressed as $a_j = [x_j, p_j]$, where $x_j$ is the subchannel allocation variable and $p_j$ is the transmitted power in the chosen subchannel.

At time slot $t$, each agent observes its current state information and determines the transmit power allocation on the available subchannels according to the outputs of the DDQN. Once a decision is made, each authorized UE switches to the assigned subchannels and adjusts the transmission power.

\subsection{Multi Agent SAPA-DDQN}
\begin{figure*}
    \centerline{\includegraphics[width=0.4\textwidth]{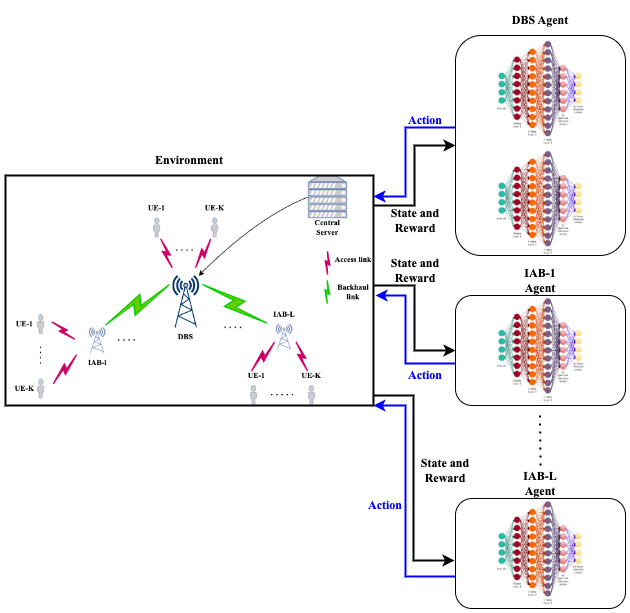}}
    \caption{Multi Agent Reinforcement Learning in an IAB Network}
    \label{marl}
\end{figure*}
Note that the optimal policy $\pi_j^*: S \longrightarrow A_j$ is obtained to maximize the long-term reward in single-agent reinforcement learning. Considering its simplicity and distributed characteristics, collaborative multi-agent reinforcement learning is
considered with local states. More precisely, each BS tries
to learn the optimal policy ${\pi_j}^*$ and learns to take the best action ${a_j}^*\in A_j$ based on its network state.

Since we do not have a central controller which collects information on the overall network, we save on transmission overheads and also keep the information associated with each agent in the network confidential. The agents interact with the environment in a decentralized manner, but the cumulative reward is calculated for the cumulative network in a centralized fashion. In our case, we have two types of state spaces, state spaces for the DBS and IABs associated with its UEs and state space for the DBS with its associated IAB nodes. For readability, we split the former two into the DBS-UE-associated state space and IAB-UE-associated state space.

As mentioned earlier, the state space vector associated with each of the agents is a one-hot encoded vector which is activated only when the associated user rate is greater than the set threshold. That is, for each user group associated with the DBS and IABs, the corresponding cell for which the data rate is grater than $R_{th}$ is activated.
\begin{equation}
    \vec{s}_{l,t}(k) = 
    \begin{cases}
        1 & \text{if}\ R_{k_{l}} \geq R_{th}\\
        0 & \text{otherwise}
    \end{cases}
\end{equation}
    
The user rate is determined using the interference and channel state information from the IAB node to the UE. This rate is calculated using the equations \eqref{2} - \eqref{7}. The state space for each agent changes at every timestamp $t$ as the agent learns to adopt the optimal policy.

Each action space corresponding to an agent contains information about the subchannel and allocated power (a random bivariate vector with finite discrete sample space). The action space is the output we receive from the DDQN network associated with each of the agents in the MADRL framework of our problem.

$R_t$ is the future cumulative discounted reward at time $t$ and is expressed as

\begin{equation}
    R_t = \sum^{\infty}_{\tau = 0}\gamma^{\tau}r_{t+\tau+1}
\end{equation}
where $\gamma \in [0,1]$ is a discounting factor for the future reward. Since we want the policy to maximize both the long-term and short-term rewards, the main aim of the Multi-Agent framework is to find such a $\pi$ that allows this. To elaborate, as $\gamma \xrightarrow{} 1$, the agent is more strongly influenced by the future rewards than the immediate ones. As $\gamma \xrightarrow{} 0$, the agent takes the short term reward more strongly, hence can be termed as myopic.

The agent learns to find a policy that treads finely between this trade-off through the Q-learning algorithm expressed by the Q-function for a given state-action pair $(s, a)$, which is defined as
\begin{equation}
    Q^{\pi}(s, a) = \mathbb{E}_{\pi}[R_t|s^t=s,a^t=a]
\end{equation}
where $\mathbb{E}_{\pi}[.|.]$ refers to conditional expectation.

To improve the policy of each agent, we greedily choose the action for the state space at time step $t$ as follows
\begin{equation}
    a^t = \arg \max_{a \in \mathcal{A}} Q(s^t, a)
\end{equation}
The action dictates the next state the agent will take, hence the Q-function will be updated accordingly. The Q-update we use in our problem is as follows
\begin{equation}
    Q_{new}(s^t, a^t) = Q_{old}(s^t, a^t) + 
\end{equation}
\begin{equation*}
    \alpha\underbrace{[r_t + \beta \max_{a \in \mathcal{A}} Q_{old}(s^t, a) - Q_{old}(s^t, a^t)]}_{TD\ Error}
\end{equation*}
here, $\alpha$ is the learning rate which determines how quickly we want the agent to learn. When $\alpha$ is too small, the model may take a very long time to converge to the optimum, whereas when it is large, it may land at a sub-optimal solution. As the agent approaches the optimal policy $\pi^*$, this updated value becomes negligible. This updated value is termed the \textit{temporal} difference error since the update rule is based on the estimated error between two different time stamps.

Since we are using a Deep Neural Network for approximating the Q-function, we update the network weights after each time step through a back-propagation step which in turn uses an appropriate loss function. We make use of \textit{smooth-l1-loss} as it prevents explosive gradients and is less sensitive to outliers compared to \textit{MSE-loss}. The DQN updates the weights of the Q-function, hence the Q-Table is approximated by the DQN as follows
\begin{equation}
    \hat{Q}_{DQN}(s, a; \theta) \approx Q(s, a)
\end{equation}
Here, $\theta$ refers to the weights of the DNN used to approximate the Q values.
In the case of a DDQN, we have two networks, a train network and a target network. The main aim of the DQN is to improve and stabilize the training procedure of Q-learning. However, in many cases, this can lead to overfitting and over-optimistic value estimates. Like Q-learning, a Double Deep Q-Network model collects and stores the information of the state-action pairs of the agents interacting with the environment in an \textit{experience replay buffer}. For every time step $t$, $(s^t_j, a^t_j, r_t, \hat{s}^t_j)$ is stored in a buffer $B$. Random sampling helps tackle problems related to correlated data and non-stationary distributions. The train DQN updates its weights $\theta$ by randomly sampling a batch of size $b$ from the buffer $B$ at each time step $t$. The target network, whose weights are a stale copy of the train network, is updated periodically according to the hyperparameter settings in the model. The weights of the target network are denoted by $\theta^{-}$. This target network is referred to as \textit{quasi-static} due to the periodic updation. Here the loss function is modified to suit the incorporation of the target network. In a DDQN the train Q-network is used to select the actions and the target Q-network is used to evaluate the loss over the chosen action. The target Q-value is calculated as
\begin{equation}
    y_{DDQN} = r_t + \gamma \hat{Q}_{DQN}\left(\tilde{s}, \arg \max_{a' \in \mathcal{A}} \hat{Q}_{DQN}(s', a';\theta);\theta^{-}\right)
\end{equation}
This optimization problem is solved using the \textit{RMSprop} optimizer to minimize the loss. The loss function is described as
\begin{equation}
    \mathcal{L}(\theta) = \arg \min_{\theta} 
    \begin{cases}
    \sum^{b}_{i} \frac{1}{2}\alpha^2 &\text{, for $\alpha < \delta$}\\
    \sum^{b}_{i} \delta(|\alpha|-\frac{1}{2}\delta) &\text{, otherwise}
    \end{cases}
\end{equation}
where $\alpha = y_{DDQN} - \hat{Q}_{DQN}(s,a;\theta)$.

The DDQN is trained in a decentralized manner, however, the reward received by the agents is the average of the individual rewards the log-sum rate of UEs associated with the entire network, which retains the notion of centralization. Since the DBS has IAB nodes and UEs associated with it, we allocate two DDQN networks associated with the DBS - one which allocates for the IABs and for the UEs associated with the DBS. This idea ensures we are dealing with a smaller state and action space, leading to lesser memory usage and faster computation.

The network demonstrates its capabilities by learning an optimal policy within a few episodes of training. We also add random perturbations in the UE positions as in real-life scenarios it is not possible to have UEs stationary. In the next results section, we record and show the results obtained in both cases.

\section{SIMULATION RESULTS}\label{sim}
This section demonstrates the simulation results obtained using our algorithm and the baselines for efficient spectrum allocation in a backhaul network in different scenarios as outlined below.

\subsection{Simulation Setup}


\begin{table}[]
    \centering
    \begin{tabular}{|c|c|}
    \hline
        Parameter & Value \\
    \hline
        Carrier Frequency & $2.4$ GHz \\
        Bandwidth $W$ & $200$ MHz \\
        DBS Pathloss & $34 + 40 \log(d)$ \\
        IAB Pathloss & $37 + 30 \log(d)$ \\
        Subchannel & Rayleigh Fading \\
        Transmit Power at DBS & $46\ dBm$ \\
        Transmit Power at IAB & $33\ dBm$ \\
        Self-interference & $-70\ dB$ \\
        Spectral Noise & $-174\ dBm$/Hz$ + 10\log(W) + 10\ dB$ \\
    \hline
    \end{tabular}
    \vspace{2mm}
    \caption{SIMULATOR SETUP}
    \label{tab:setup}
\end{table}

We consider a network with a central DBS, associated with $L$ IAB nodes. Each of the IABs and DBS is associated with $K$ UEs. The number of antennas for the DBS, the IAB node, and UE is 64, 8, and 1, respectively. Information regarding other network parameters is elaborated in the table \ref{tab:setup}. We adopt 5GCM UMa for the channel fading model, which is in accordance with 3GPP standards\cite{Inoue20205GNR}. The weighting factors for the rewards for each of the agents are set so that we obtain the average of the rewards over all the users present in the network considered. A penalty of $500$ is also added, which penalizes the agent for making a sub-optimal decision. We now elaborate on the hyperparameters and network model used to achieve the solution to the resource allocation problem. The DDQN in the proposed scheme employs the network structure consisting of three hidden layers: fully-connected neural networks with 200, 400, and 800 neurons, respectively. The discount factor $\gamma = 0.95$ and the learning rate $\alpha=10^{-3}$. We consider a buffer size of $5000$ for the \textit{experience replay buffer} $B$ and a batch size of $b=16$ for the random sampling of the buffer. 

We compare our model against the \textit{DQN} and \textit{Random} policies as the baselines to evaluate our model. As seen from the simulation visualizations, our model outperforms the \textit{Random} and \textit{DQN} methods by a large margin. Since the \textit{Random} method samples random actions without taking the efficient allocation of subchannels and power into account, the system is exposed to a lot of interference which leads to a degradation in the sum rate of the UEs. The state and action spaces in the DQN explode with the increase in IAB nodes and UEs, hence it takes a lot of time to train.

The \textit{SAPA-DDQN} effectively tackles this problem by exploring possible state-action pairs to allocate spectrum and power efficiently. In the case of the \textit{DQN} model, training can lead to over-fitting, which is avoided in the DDQN architecture. Further, to optimize our model's available action space size, we employ two DDQN networks for agents associated with the DBS, one for predicting a suitable action for the UEs associated with the DBS, and another for choosing an optimal action for the IABs associated with the DBS. The final action is a combination of the two decisions taken by independent DDQNs. This is used in a unified fashion to calculate the reward associated with the DBS. This method is more effective than using a single DDQN associated with the DBS.

At every time step $t$ of the simulated iterations, we move the user to a position $(x_t, y_t) = (x_0, y_0) + \mathcal{U}(0, 2)\times(cos(\mathcal{U}(0, 2\pi)), sin(\mathcal{U}(0, 2\pi)))$ at each time step $t$, where $\mathcal{U}(a, b)$ denotes a uniform distribution generator between the range $(a, b)$. In our case since the radius around which the UE is allowed to perturbate is $2m$, $a=0, b=2$, $(x_0, y_0)$ is the initial position of the UE and $(x_t, y_t)$ is its position at time step $t$.

\subsection{Results}
In this subsection, we present the results of all our simulations of the Q-network.

We consider two main cases - one where the users are in a constant position and another where their positions are perturbed. In both cases, we compare the results for similar networks. Since the state and action spaces blow up exponentially with the increase in the number of subchannels, users, and IABs, we present the results for smaller networks. This case is computationally less intensive and requires lesser memory. Since the DBS has two DDQNs associated with it, the combinations for the state and action space for the UEs and IABs associated with the DBS require exponentially lesser memory as they are split between two DDQN networks. Each episode runs for 1000 steps in all our simulations. Smaller IAB topologies converge faster than larger ones.

\begin{figure}[h]
    \centering
        \includegraphics[scale=0.30]{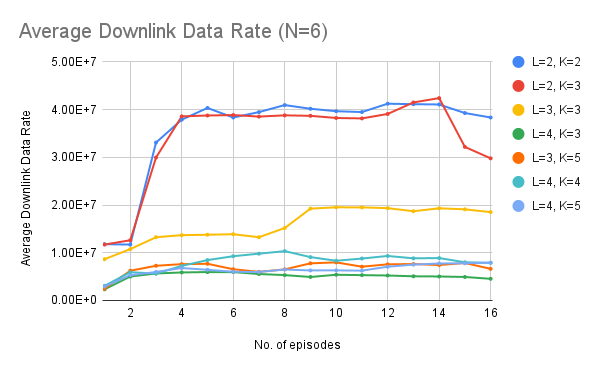}
    \caption{Graph demonstrating the average downlink data rate for different configurations}
    \label{fig3}
\end{figure}

\begin{figure}[h]
    \centering
        \includegraphics[scale=0.30]{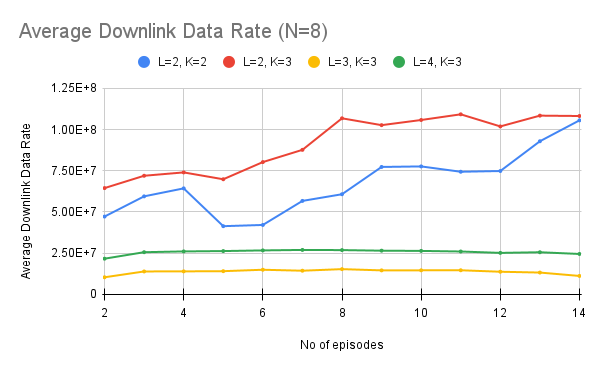}
    \caption{Graph demonstrating the average downlink data rate for different configurations}
    \label{fig4}
\end{figure}

Figure \ref{fig3} shows the graph illustrating the average downlink rate among the multiple users for different configuration setups according to $L$ (number of IABs) and $K$ (number of users) for $N = 6$. It can be noticed that there is a steep rise in the downlink data rate after training for a number of episodes after which the data rate converges to a stable value. Similarly, Figure \ref{fig4} demonstrates the same for $N = 8$ subchannels. This also converges to a stable value in the long run but takes a long time to execute due to higher computational complexities.

The other set of experiments we carried out is for perturbed user positions. We consider a system of $L=3$, $K=3$, and $N=6$. It can be observed that these graphs peak a few episodes into training and settle back to the initial downlink data rate. Since the positions of the users are changing rapidly for every step of each episode, the model is unable to learn a correct representation of weights for the DDQN.

\begin{figure}[h]
    \centering
        \includegraphics[scale=0.30]{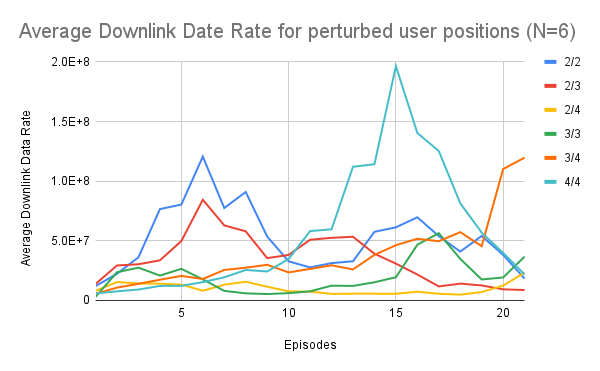}
    \caption{Graph demonstrating the average downlink data rate for different configurations in the case of moving users}
    \label{fig5}
\end{figure}

\begin{figure}[h]
    \centering
        \includegraphics[scale=0.30]{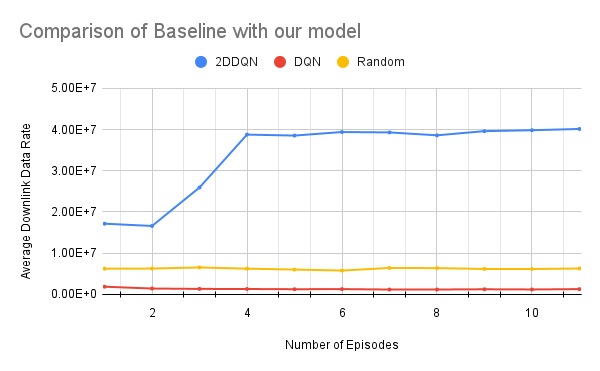}
    \caption{Comparison of our model with other models}
    \label{fig6}
\end{figure}

Finally, Figure \ref{fig6} shows how the model fares against the baselines for $L=3, K=3$, and $N=8$. Here it is seen that our \textit{SAPA-DDQN} model performs almost 10-12$\times$ better than the \textit{DQN} and \textit{Random} models.

\section{CONCLUSION}\label{conc}
In this paper, we propose the \textit{SAPA-DDQN} method for training a model based on simulating our custom environment to solve the optimization problem regarding resource allocation in an IAB Network. We successfully demonstrate that our algorithm performs better than the baselines and can be trained offline with little to no numerical information about the network at hand. The simulation results for perturbed user positions also give promising results for maximizing the average downlink data rate among the users. \textit{SAPA-DDQN} model performs almost 10-12$\times$ better than the \textit{DQN} and \textit{Random} models. We also find that the model converges very quickly for smaller networks. Future work directions would be understanding how networks with more users and associated IAB nodes can be handled.

\bibliographystyle{plain}
\bibliography{bibliography.bib}
\end{document}